\documentclass[]{./aa}    
\usepackage{graphics}

\begin{document}
\thesaurus{07         
               07.09.1;  
               13.09.1;  
               12.12.1;  
               04.03.1;  
               04.19.1;  
               09.04.1;  
             }

\title {Calibration of the distance scale from galactic Cepheids:I}
\subtitle{Calibration based on the GFG sample. }
\titlerunning{Calibration of the distance scale  }

\author{Paturel G. \inst{1}, Theureau G. \inst{4}, Fouqu\'e P. \inst{2},
Terry J.N. \inst{1}, Musella I. \inst{3}, Ekholm T. \inst{1}}

\authorrunning{Paturel G., et al.}
\offprints{G. Paturel - 
The table of the Appendix and Table 2 are available in electronic form
at CDS and on our anonymous ftp-server www-obs.univ-lyon1.fr (pub/base/CEPHEIDES.tar.gz).}

\institute{
              CRAL-Observatoire de Lyon,\\
              Avenue Charles-Andr\'e
              F69561 Saint-Genis Laval CEDEX, FRANCE \\
\and
              European Southern Observatory \\
              Casilla 19001      \\
              19 Santiago, CHILE  \\
\and
              Osservatorio Astronomico di Capodimonte \\
              Via Moiariello 16, \\
              80131 Napoli, ITALY \\
\and
              Laboratoire de Physique et de Chimie de l'Environnement \\
              3A Avenue de la Recherche scientifique   \\
              45071 Orleans cedex 02, FRANCE  \\
}

   \date{Received 6 March 2001 / Accepted 28 November 2001 }
 
   \maketitle

   \begin{abstract}
New estimates of the distances of 36 nearby galaxies are presented
based on accurate distances of galactic Cepheids obtained by
Gieren, Fouqu\'e and Gomez (1998) from the  geometrical
Barnes-Evans method.  

The concept of 'sosie' is applied to extend the 
distance determination to extragalactic Cepheids without 
assuming the linearity of the PL relation. 
Doing so, the distance moduli are obtained in a straightforward way.

The correction for extinction is made using two photometric
bands ($V$ and $I$) according to the principles introduced by
Freedman and Madore (1990).
Finally, the statistical bias due to the incompleteness of the
sample is corrected according to the precepts
introduced by Teerikorpi (1987) without introducing any free parameters 
(except the distance modulus itself in an iterative scheme).

The final distance moduli depend on the adopted extinction 
ratio ${R_V}/{R_I}$ and on the limiting apparent magnitude of
the sample.
A comparison with the distance moduli recently published 
by the Hubble Space Telescope Key Project (HSTKP) team reveals a 
fair agreement when the same ratio ${R_V}/{R_I}$ is used but
shows a small discrepancy at large distance. 

In order to bypass the uncertainty due to the metallicity effect it
is suggested to consider only galaxies having nearly the same 
metallicity as the calibrating Cepheids (i.e. Solar metallicity).
The internal uncertainty of the distances is about 0.1 magnitude 
but the total uncertainty may reach 0.3 magnitude.
   \end{abstract}
\section{Introduction: Discussion of the problems related to Cepheids}
As an extension of our study of the kinematics of the local universe
(KLUN+) we need an accurate value for the global Hubble constant and
accurate distances of individual galaxies. The calibration of the
distance scale is thus a fundamental step in this process.
The aim of this work was to calibrate
the distance scale from nearby galactic Cepheids for which the 
HIPPARCOS satellite measured geometrical parallaxes.
This should avoid the step of calibrating the distance scale by 
assuming a given distance to the Large Magellanic Cloud (LMC).
Unfortunatelly, it turns out that these measurements are very difficult
to use due to a statistical bias (Lutz and Kelker, 1973). 
The difficulties can be solved by proper treatment, like the one 
proposed by Feast and Catchpole (1997). It has been shown that this leads 
to unbiased results (Pont et al., 1997; Lanoix et al. 1999), 

On the other hand, individual measurements of Cepheids from HIPPARCOS
are relatively inaccurate because of the distance of galactic Cepheids.  
Excluding $\alpha$ UMi which does not pulsate in the
fundamental mode, the best geometrical parallax of an individual Cepheid 
obtained from HIPPARCOS is 3.32$\pm$ 0.58 marcsec for $\delta$ Cephee.
This leads to an uncertainty in the distance modulus of 0.38 magnitude. 
In comparison, the {\it quasi-geometrical} method of Barnes-Evans applied 
to Cepheids (Gieren, Fouqu\'e, Gomez, 1998; hereafter GFG), 
gives distance moduli with a typical uncertainty less than 0.1 magnitude
(the external error can be estimated to about 0.2 magnitude according to 
Table 7 in GFG). 
We call this method {\it quasi-geometrical} because it requires only a few
assumptions.
The method is independent of any determination of the LMC distance
and has a relatively small systematic error (about 0.2 magnitude).
{\it Thus, we decided to calibrate the distance scale using the work done
by Gieren, Fouqu\'e and Gomez (1998).}

Nevertheless, other difficulties appear. The slope of the 
Period-Luminosity relation (hereafter, PL relation) determined from
the adopted calibrating galactic Cepheids differs from the slope obtained 
for the LMC by the same authors (GFG) (Table \ref{slopes})..
For the LMC, the slopes in V and I bands are now confirmed by the 
OGLE survey (Udalski. et al., 1999). 
What slope should we adopt?
 
\begin{table}
\caption{Slopes of the PL relation.}
\label{slopes}
\begin{tabular}{lll}
\hline
source & $a_V$ &  $a_I$  \\
\hline
GFG(MW)  & $-3.037\pm 0.138$  & $-3.329 \pm 0.132$  \\
GFG(LMC)  & $-2.769\pm 0.073$  & $-3.041 \pm 0.054$  \\
OGLE(LMC) & $-2.765$  &$-2.963$ \\
\hline
\end{tabular}
\end{table}

The true physical relation is actually a Period-Luminosity-Color 
(hereafter, PLC) relation written as 
$M=\alpha \log P + \beta C_o + \gamma$, where $M$ is
the absolute magnitude and $C_o$ the intrinsic color.
The PL relation is simply the projection of the PLC onto the P-L plane. 
In the PLC relation the slope $\partial M / \partial logP$ is constant.
However, the observed slope of the PL relation depends on the
distribution of observed Cepheids in the PLC plane (i.e., on the color 
distribution of the sample). Hence, the slope in a given photometric band
may partially depend on the metallicity, because it affects the intrinsic
color. 
Linear non-adiabatic models do predict that the slope is constant when one 
uses bolometric magnitudes (Baraffe et al., private communication),
whereas non-linear models predict that the slope depends on
the metallicity also for the bolometric magnitudes
(Bono et al., 2000 and references therein) and
predict that the slope in a given band depends on the metallicity.
Because the metallicity of the LMC differs from the metallicity in
the Solar neighbourhood, the choice of slopes in different
bands is difficult. 
{\it In order to avoid this dilemma we decided to apply the method
of 'sosie' (Paturel, 1984) because it does not require knowledge
of the slope and zero point of the PL relation
\footnote{this method was first introduced to solve the same
kind of problems for the Tully-Fisher relation (1977)}.}

The correction for extinction produced by interstellar matter
is another difficulty. It can
be solved by assuming that the extinction law is universal. 
We will thus assume that the extinction on an apparent magnitude
is proportional to the color excess 
($A_{\lambda} = R_{\lambda} (C-C_o)$, where $C$ is the reddened color). 
The factor of proportionality $R_{\lambda}$ is taken from
tabulations (e.g., Cardelli, Clayton \& Mathis, 1989 ; 
Caldwell \& Coulson, 1987 ; Laney \& Stobie, 1993). 
It depends on both the considered band and color.
With such an assumption it is possible to use
the Freedman and Madore (1990) precepts of de-reddening.
Two bands are needed in order to calculate a color. Because
most extragalactic Cepheids are measured in V- and I-band from
The {\it Hubble Space Telescope} (hereafter, HST), we will use these
two bands.
{\it Thus, the Freedman and Madore (1990) de-reddening method will be adapted 
to the sosie method, used in V and I photometric bands.}

Finally, an ultimate difficulty comes from the incompleteness bias.
This bias was first studied by Teerikorpi (1987) for application
to galaxy clusters (Bottinelli et al., 1987). 
It was first denounced by Sandage (1988) in application to
the PL relation and re-discussed later by 
Lanoix, Paturel and Garnier (1999a). The sample to which we are
applying the PL relation must be statistically representative of 
the calibrators themselves.  
Indeed, due to the intrinsic scatter of the PL relation,
there is a given distribution of absolute magnitudes at a given period. 
At increasing distances the fainter end of this distribution is
progressively missed and the distribution of the actual sample changes. 
Restricting the sample to Cepheids with a period larger than a given limiting 
period reduces this bias. 
The limiting period depends on a first estimate of the distance, 
on the apparent limiting magnitude
and on the characteristics of the PL relation (dispersion, slope and 
zero-point). In fact, the full theory of Teerikorpi is applicable. 
The method is much more complete than the rough
rule of thumb used as a quick approach in an application in which a detailed
treatment was not needed. However, we want to derive final distance moduli
and the precise bias correction must be used.
Note that the slope and zero point of the PL relation
are needed but only as second order terms and thus, the uncertainties 
mentioned about their choice do not present any significant 
difficulty (this will be confirmed in section 4.3).
{\it The incompleteness bias will be corrected using the precepts
given by Teerikorpi (1987).}

In section 2 we will describe the material used for this study:
the calibrating sample by GFG and our extragalactic
Cepheid database (Lanoix et al., 1999b).

In section 3 we describe the 'sosie' method  and give the basic 
equation for the calculation of the distance modulus of an
extragalactic Cepheid.

In section 4 we give the results obtained for 1840 Cepheids belonging
to 36 nearby galaxies described in the previous section.
We also discuss these results and compare them with those recently
published by Freedman et al. (2001).

\section{Observational material }
The guideline in the constitution of the observational
material is the selection of the most secure observations.
This leads us to reject some data, as explained below, both galactic 
and extragalactic.

\subsection{The list of galactic Cepheids}
The starting point of our study is the choice of the galactic
Cepheids used for the calibration. We adopt the list given
in Gieren, Fouqu\'e and Gomez (Table 3 in GFG) but we
rejected three Cepheids (EV Sct, SZ Tau and QZ Nor) because
they do not pulsate in the fundamental mode (they are overtone
Cepheids).
They correspond to the three lowest periods of the list. 
Because we use only the V and I photometric bands, three Cepheids
are also rejected (CS Vel, GY Sge and S Vul) because they do not have
I-band magnitude. Thus, 28 Cepheids remain. Their distance
moduli are adopted directly from Table 5 given by GFG.
Only three Cepheids have a mean error in their
distance modulus larger than $0.1$ magnitude.
We give in Table \ref{cali} the adopted calibrating sample
of galactic Cepheids.

\begin{table}
\caption{Adopted calibrating sample of galactic Cepheids.
{\bf Column 1:} Name of the galactic Cepheid;
{\bf Column 2:} log of the period ($P$ in days);
{\bf Column 3:} adopted distance modulus
and its mean error according to Gieren, Fouqu\'e and Gomez (1998); 
{\bf Column 4:} Mean V-band apparent magnitude;
{\bf Column 5:} Mean I-band apparent magnitude.
}
\begin{tabular}{lrrrr}
\hline
Cepheid &logP &   $\mu \pm  m.e.$& $\langle V \rangle$& $\langle I \rangle$    \\
\hline
BF Oph  &0.609&   9.50$\pm$  0.11& 7.33 &6.41\\
T Vel   &0.666&  10.09$\pm$  0.02& 8.03 &7.01\\
CV Mon  &0.731&  10.90$\pm$  0.05&10.31 &8.68\\
V Cen   &0.740&   9.30$\pm$  0.02& 6.82 &5.81\\
BB Sgr  &0.822&   9.24$\pm$  0.02& 6.93 &5.84\\
U Sgr   &0.829&   8.87$\pm$  0.01& 6.68 &5.45\\
S Nor   &0.989&   9.92$\pm$  0.03& 6.43 &5.41\\
XX Cen  &1.039&  10.85$\pm$  0.06& 7.82 &6.75\\
V340 Nor&1.053&  11.50$\pm$  0.13& 8.38 &7.15\\
UU Mus  &1.066&  12.26$\pm$  0.09& 9.78 &8.49\\
U Nor   &1.102&  10.77$\pm$  0.07& 9.23 &7.36\\
BN Pup  &1.136&  12.92$\pm$  0.05& 9.89 &8.55\\
LS Pup  &1.151&  13.73$\pm$  0.04&10.45 &9.06\\
VW Cen  &1.177&  13.01$\pm$  0.04&10.24 &8.77\\
VY Car  &1.277&  11.42$\pm$  0.04& 7.46 &6.28\\
RY Sco  &1.308&  10.47$\pm$  0.04& 8.02 &6.30\\
RZ Vel  &1.310&  11.17$\pm$  0.03& 7.09 &5.85\\
WZ Sgr  &1.339&  11.26$\pm$  0.02& 8.02 &6.53\\
WZ Car  &1.362&  12.98$\pm$  0.14& 9.26 &7.95\\
VZ Pup  &1.365&  13.55$\pm$  0.04& 9.63 &8.28\\
SW Vel  &1.370&  11.99$\pm$  0.06& 8.12 &6.83\\
T Mon   &1.432&  10.58$\pm$  0.07& 6.12 &4.98\\
RY Vel  &1.449&  12.10$\pm$  0.05& 8.37 &6.84\\
AQ Pup  &1.479&  12.75$\pm$  0.04& 8.67 &7.12\\
KN Cen  &1.532&  12.91$\pm$  0.06& 9.85 &7.99\\
$\iota$ Car&1.551&   8.94$\pm$  0.05& 3.73 &2.59\\
U Car   &1.589&  11.07$\pm$  0.04& 6.28 &5.05\\
SV Vul  &1.654&  12.32$\pm$  0.07& 7.24 &5.75\\
\hline
\end{tabular}
\label{cali}
\end{table}

\subsection{The list of extragalactic Cepheids}
In 1999 we have constructed an Extragalactic Cepheid database
(Lanoix et al., 1999b) by collecting 3031 photometric measurements
of 1061 Cepheids located in 33 galaxies. This list has been
updated. Especially, the V and I band measurements by Udalski et al. 
(OGLE survey, 1999) were added for the LMC from the data
available through 'astro/ph9908317'.
The new database contains 6685 measurements for 2449 Cepheids in 46
galaxies. In order to make this compilation available, 
the full contents of the extragalactic part will be published in 
electronic form for the A\&A archives at CDS. A description is given 
in the Annex. 

In this database, each light curve has been inspected in order to
describe the main features. In the present study only light curves
considered as 'Normal' are used
\footnote{Lanoix et al. give eight classes
of light curves:  'Normal', 'Symetrical', 'Bumpy', 'Scattered',
'Overtone', 'Low amplitude', 'Peculiar', 'No curve'. A 'Normal' light curve
is characterized by a non-symetrical variation: a fast increase and a slower 
decrease.}.  We reject all peculiar light curves including light curves 
classified as 'low amplitude' because they are often associated with overtone
Cepheids.

Only the mean V and I band magnitudes are kept. When several
magnitudes are averaged from different sources we keep the mean
only if the mean error is less than 0.05 magnitude.
It is to be noted that HST measurements of seven galaxies 
\footnote{IC4182, NGC3368, NGC3627, NGC4496A, NGC4536, NGC4639 and NGC5253}
have been analyzed by two independent groups. This leads to two different
sets of magnitudes. Independent treatment of both sets shows that
the distance modulus differs by less than 0.1 magnitude, except for
IC4182 for which the difference is 0.28 magnitude (Lanoix, private
communication). Because we have
no means to decide which set is the best we decided to keep them
both. 

The final catalogue (Table \ref{cati}) results in 1840 extragalactic 
Cepheids.  They belong to 36 galaxies, 27 of which come from HST observations
and 9 from ground-based observations. The full Table is available in 
electronic form in the A\&A archives at CDS.

\begin{table}
\caption{Sample of extragalactic Cepheids.
{\bf Column 1:} Name of the host galaxy;
{\bf Column 2:} Name of the Cepheid according to the following reference;
{\bf Column 3:} Reference (coded) from which the Cepheid name is taken;
{\bf Column 4:} log of the period ($P$ in days);
{\bf Column 5:} Mean V-band apparent magnitude;
{\bf Column 6:} Mean I-band apparent magnitude.
Only a part of the table is given. The rest is available in 
electronic form.
}
\begin{tabular}{lllrrr}
\hline
galaxy &Cepheid & Ref. &logP &$\langle V \rangle$&$\langle I \rangle$\\
\hline
IC4182 &   C11  &Gib99 & 1.423&23.10&22.21\\
LMC    &109838  &Uda99 & 0.732&16.14&15.11\\
NGC1365&V32     &Sil98 & 1.460&26.77&25.94\\
NGC1425&C15     &Mou99 & 1.295&26.63&25.90\\
NGC2090&C13     &Phe98 & 1.461&25.44&24.55\\
NGC224 &FI13    &Fre90 & 1.497&19.24&18.33\\
NGC2541&C25     &Fer98 & 1.270&25.68&24.90\\
NGC3031&C13     &Fre94 & 1.270&23.56&22.75\\
NGC3109&P2      &Mus98 & 0.722&22.18&21.87\\
NGC3198&   C19  &Kel99 & 1.220&26.23&25.12\\
NGC3319&C13     &Sak99 & 1.398&25.61&24.89\\
NGC3351&C25     &Gra97 & 1.207&25.77&24.49\\
NGC3368&   C09  &Gib99 & 1.483&25.13&24.11\\
NGC3621&C14     &Raw97 & 1.498&23.28&22.76\\
NGC3627&   C14  &Gib99 & 1.366&24.66&23.46\\
NGC4258&MAO14   &Mao99 & 1.330&24.65&23.88\\
NGC4321&C9      &Fer96 & 1.700&25.93&24.88\\
NGC4414&C1      &Tur98 & 1.658&25.89&24.85\\
NGC4496&   C24  &Gib99 & 1.717&25.27&24.26\\
NGC4535&C35     &Mac99 & 1.390&26.14&25.22\\
NGC4536&   C12  &Gib99 & 1.484&25.81&24.89\\
NGC4548&   C09  &Gra99 & 1.270&25.96&25.38\\
NGC4603&2984    &New99 & 1.570&27.19&26.37\\
NGC4639&   C14  &Gib99 & 1.717&26.33&25.28\\
NGC4725&C09     &Gib98 & 1.590&24.85&23.87\\
NGC5253&   C07  &Gib99 & 1.025&23.71&22.86\\
NGC5457&V4      &Kel96 & 1.471&23.51&22.78\\
NGC598 &V31     &Chr87 & 1.572&19.17&18.14\\
NGC7331&V4      &Hug98 & 1.354&26.13&24.93\\
NGC925 &V18     &Sil96 & 1.439&24.99&23.97\\
SEXB   &V2      &Sa85b & 1.444&20.60&20.00\\
.... & & & &\\
.... & & & &\\
.... & & & &\\
\hline
\end{tabular}
\label{cati}
\end{table}

\section{Method of sosie}
The method of 'sosie' was introduced (Paturel, 1984) to avoid
the problem encountered in the practical use of the Tully-Fisher
relation (Tully and Fisher, 1977), a linear relation between 
the absolute magnitude
of a galaxy and its 21-cm line width. Here we are in similar
conditions with a linear relationship between the absolute
magnitude and an observable parameter, the logarithm of the period.
In French, the word 'sosie' refers to someone who looks very similar
to someone else without being necessarily genetically related.
Here two Cepheids will be considered as 'sosie' if their light curves
have the same shape and if they have the same period (within a given 
error). Because of the selection based on the shape of the light curve
we will consider that all Cepheids of our sample pulsate in the
fundamental mode. They all obey the same P-L relation.

We write the distance modulus of a calibrating Cepheid and of 
an extragalactic Cepheid through a universal PL relation. The calibrating 
Cepheid is identified with subscripts 'c' and no subscript for the
extragalactic one. 
Presently, we assume that both stars have the
same metallicity and the same intrinsic color. We will see
how to bypass this problem, later. 

\begin{equation}
\mu_c = m_{c}^{o} - a \log P_c - b,
\end{equation}
\begin{equation}
\mu = m^{o} - a \log P - b
\end{equation}
$m^{o}$ is the apparent mean magnitude in a given band. 
The superscript '$o$' means 'corrected for extinction'.
If one selects an extragalactic Cepheid having the same period
as the calibrating one, i.e., $\log P = \log P_c$, the
distance modulus of the extragalactic Cepheid is then
\begin{equation}
\label{eqmu}
\mu = \mu_c + m^{o} - m^{o}_{c}
\end{equation}
The distance modulus of the extragalactic Cepheid is deduced
without having to know the slope and zero-point of the PL relation.

In order to correct for extinction we apply the previous equation
to two different bands and express the extinction term
as a function of the color excess $E = E(B-V)$. In order to make the
notations clearer we note the apparent magnitudes 
$V$ and $I$ for the two considered bands. From Eq. \ref{eqmu} one has:
\begin{equation}
\mu = \mu_c + V - R_V E - V_{c} +  R_V E_{c}
\end{equation}
\begin{equation}
\mu = \mu_c + I - R_I E - I_{c} +  R_I E_{c}
\end{equation}
which can be written as
\begin{equation}
\mu = \mu_c + V - V_{c} - R_V ( E -  E_{c} )
\end{equation}
\begin{equation}
\mu = \mu_c + I - I_{c} - R_I ( E -  E_{c} )
\end{equation}
Then, eliminating $ E -  E_{c}$ between the two previous equations
we obtain:

\begin{equation}
\mu = \mu_c + \frac {(V - V_{c}) - \frac{R_V}{R_I} (I - I_{c})}{ 1 - \frac{R_V}{R_I} }
\end{equation}
This is the desired equation. It can be written in a more elegant
manner by using the reddening-free Wesenheit function (Van den Bergh, 1968):
\begin{equation}
W = \frac{V - (R_V / R_I). I}{1 - (R_V / R_I)}
\end{equation}
\begin{equation}
\mu = \mu_c + W - W_c 
\end{equation}
In practice, the intrinsic color is not known and this equation is valid
only for Cepheids of the same intrinsic color and metallicity. 
Thus, for a true sample, we will write it as (see the discussion below):
\begin{equation}
\label{kappa}
\langle \mu \rangle = \mu_c + \langle W - W_c \rangle
\end{equation}

$W$ is an observable quantity. Then, the mean distance
modulus of a sample of Cepheids which have the same period of 
pulsation as a calibrating Cepheid can be obtained directly from 
Eq. \ref{kappa}.

The physical relationship in this result is a
Period-Luminosity-Color relation. This means that we should search for
sosie of calibrators by considering both their similarity in $\log P$
and intrinsic color $C_o$. But the intrinsic color is not observable.
Thus, equation \ref{kappa} must be considered as a statistical relation
exactly as the PL relation. Because of the statistical 
relation between $C_o$
and $\log P$, the selection in $\log P$ will guarantee that a
calibrator Cepheid and its sosies have, {\it on average}, 
the same intrinsic color.
So, the problem of the intrinsic color is partially bypassed.
For the metallicity problem, the solution is to consider that
the method is valid only for galaxies having nearly the same metallicity
as the calibrating Cepheids. In the present paper this means that,
{\it stricto sensu}, only galaxies with a nearly Solar metallicity 
can be considered as valid. In practice, we applied the method to different
kinds of galaxies without noting strong metallicity dependence.

As a test, we apply the method to the calibrating sample itself. Indeed, 
some galaxies of the sample can be considered as sosie of another. Note that
each calibrating Cepheid has at least itself as a sosie. Obviously,
we will not consider this special case.
We will accept two Cepheids as sosie when the difference of their
$\log P$ is smaller than $0.07$. With a PL slope of $\approx -3$, this
will give an uncertainty $\approx 0.2$ mag. in the distance modulus.
We adopt the ratio ${R_V}/{R_I}=1.69$ because it corresponds to
the most widely accepted one (it corresponds to a ratio of total-to-selective 
absorption $A_V / (A_V - A_I)= R_V / (R_V - R_I) = 2.45$).

\begin{figure}[!]
\resizebox{\hsize}{!}{\includegraphics*{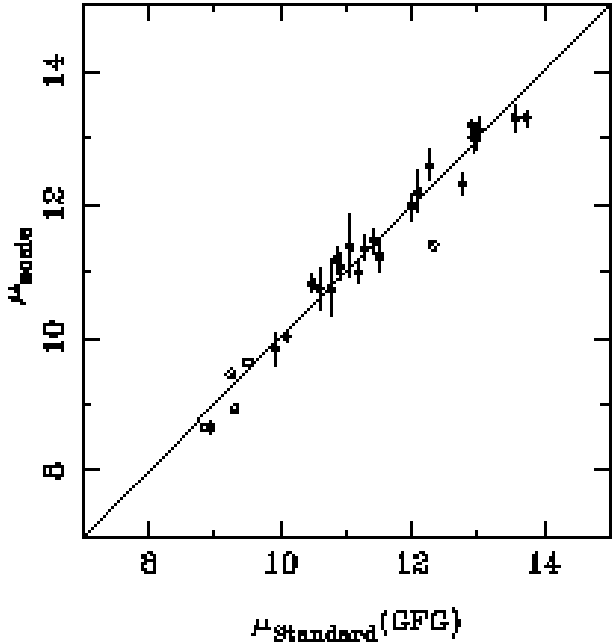}}
\caption{Comparison of standard distance moduli with those calculated
from the method of sosie.
The solid line corresponds to a slope of one and a zero-point of zero. 
Open circles represent
the points for which there is only one determination and then no
standard deviation.
}
\label{fig001}
\end{figure}

In table \ref{calisos} we give the distance moduli obtained with
equation \ref{kappa} for 23 Cepheids which are sosie of another 
calibrator. In Figure \ref{fig001} the comparison of the calculated
distance moduli with the calibrating ones is given.

\begin{table}
\caption{Distance moduli we obtain with
equation \ref{kappa} for 28 Cepheids which are sosie of another
calibrator. 
{\bf Column 1:} Name of the galactic Cepheid;
{\bf Column 3:} Distance modulus calculated from Eq.\ref{kappa};
{\bf Column 3:} Standard deviation on the calculated distance modulus;
{\bf Column 5:} Number of sosies (excluding its own case).
}
\begin{tabular}{lrrr}
\hline
Cepheid  &  $\mu$&  std.dev.& n  \\
\hline
T Vel   &   10.03&    0.08&    2\\
BF Oph  &    9.63&     -  &    1\\
CV Mon  &   11.07&    0.20&    2\\
V Cen   &    8.93&     -  &    1\\
U Sgr   &    8.65&     -  &    1\\
BB Sgr  &    9.46&     -  &    1\\
XX Cen  &   11.17&    0.20&    4\\
V340 Nor&   11.22&    0.21&    4\\
S Nor   &    9.82&    0.24&    2\\
UU Mus  &   12.59&    0.22&    3\\
U Nor   &   10.73&    0.40&    5\\
BN Pup  &   13.00&    0.23&    3\\
LS Pup  &   13.31&    0.11&    3\\
VW Cen  &   13.14&    0.19&    2\\
RY Sco  &   10.81&    0.13&    6\\
RZ Vel  &   10.98&    0.17&    6\\
WZ Sgr  &   11.34&    0.18&    6\\
VY Car  &   11.46&    0.19&    3\\
WZ Car  &   13.04&    0.20&    5\\
VZ Pup  &   13.30&    0.19&    6\\
SW Vel  &   11.97&    0.21&    6\\
T Mon   &   10.73&    0.32&    4\\
RY Vel  &   12.19&    0.31&    2\\
AQ Pup  &   12.32&    0.15&    3\\
KN Cen  &   13.19&    0.08&    3\\
$\iota$ Car&    8.65&    0.09&    2\\
U Car   &   11.38&    0.46&    3\\
SV Vul  &   11.39&     -  &    1\\
\hline
\end{tabular}
\label{calisos}
\end{table}

From a direct regression we find that the slope is not different from one 
($1.00 \pm 0.03$). 
The observed mean difference between the calculated distance modulus and its 
standard value is obtained together with its standard deviation:

\begin{equation}
\langle  \mu_{sosie} - \mu_{standard} \rangle  = -0.00 \pm 0.24.
\end{equation}
 
The method does not introduce any systematic shift in the zero point. 
{\it This means that the calibrating Cepheids constitute
a coherent system} (at least for the 28 Cepheids used in the test).
The observed standard deviation (0.24) is in agrement with the expected
standard deviation 0.2.

\section{Application to extragalactic Cepheids}

\subsection{Preliminary determination of extragalactic distance moduli}
The method is applied to the 1840 Cepheids of Table \ref{cati}.
To accept two Cepheids as sosie, we still adopt the criterion 
$|\log P -\log P_c | < 0.07$ which guarantees that the standard deviation
is about $0.2$ mag., assuming a PL slope of $-3$.
We adopt the ratio ${R_V}/{R_I}=1.69$ which corresponds to the first order
terms proposed by Caldwell \& Coulson (1987) and Laney \& Stobie (1993).
This is also the value adopted by Freedman et al. (2001), 
following Cardelli et al. (1987), for their
HST key project about Cepheids \footnote{This value corresponds to
a ratio of total-to-selective absorption 
$A_V / (A_V - A_I)= R_V / (R_V - R_I) = 2.45$.}.
For each of the 36 host galaxies we plot the different distance moduli
given by Eq. \ref{kappa} as a function of $\log P$. 
This result appears in Figure \ref{fig003}.

The most important feature to point out is a significant trend
leading to higher distance moduli
for long period Cepheids. This trend is visible for almost all the
host galaxies. 
This is visible even for nearby galaxies if short periods
are observed. For distant galaxies the trend is visible also
at long periods.
This was expected from the 
incompleteness bias we discussed elsewhere (e.g., Lanoix et al., 1999a).
Another signature of the bias comes from the fact that only nearby
galaxies (IC1613, IC4182, LMC, NGC224, NGC3109; NGC5253) have Cepheids 
with short periods. This clearly depends on the limiting 
magnitude of the considered host galaxy.
This important question is discussed in the following subsection.

\subsection{Correction for the incompleteness bias}
In order to get the proper distance moduli we have to correct for
the incompleteness bias. 
In a previous paper (Lanoix et al. 1999a) we 
suggested using a rule of thumb to avoid this bias.
The rule consists of using only $logP$ values larger than a given 
limit $logP_l$.  This limit is expressed as:
\begin{equation}
logP_l= \frac{ V_{lim} - \mu  - b - 2\sigma}{a}
\end{equation}
Unfortunatelly, this method does not take into 
account the pieces of information contained in smaller periods. 
The detailed theory of this incompleteness bias was given
by Terrikorpi (1987) in the study of galaxy clusters. 
The bias for extragalactic Cepheids is of the same nature
because the Cepheids of a given galaxy are all at the same 
distance from us, like the galaxies of a cluster. 
Assuming that the dispersion $\sigma$ at a given $\log P$ is constant,
the basic equations adapted to the problem of
extragalactic Cepheids are the following (for the sake of simplicity
we will consider only the V band):

The observed distance modulus $\mu$ will appear smaller than the
true one. The bias $\Delta \mu$ at a given $\log P$ is:

\begin{equation}
\Delta \mu=  - \sigma {\sqrt{\frac{2}{\pi}} \frac {e^{-A^2}}{1+erf(A)}}
\label{deltamu}
\end{equation}

where

\begin{equation}
A= \frac {V_{lim} -\mu - a_v \log P - b_v} {\sigma \sqrt{2}}
\end{equation}

and

\begin{equation}
erf(x) = \frac{2}{\sqrt{\pi}} \int_0^x e^{-t^2} dt.
\end{equation}

 In these equations $a_v$ and $b_v$ are the slope and zero point
 of the PL relation. We adopt the values found by GFG from their
 galactic sample, i.e., $a_v= -3.037$ and $b_v=-1.021$
 \footnote{GFG give $b_v=-4.058$ because they consider the zero-point
 at $\log P =1$}. Note that this requirement seems to reduce the
 interest of the sosie method because the slope and zero-point are
 needed anyway. In fact, the slope and zero-point appear only as
 parameters in a second order correction.

\begin{figure*}[p]
\resizebox{\hsize}{!}{\includegraphics*{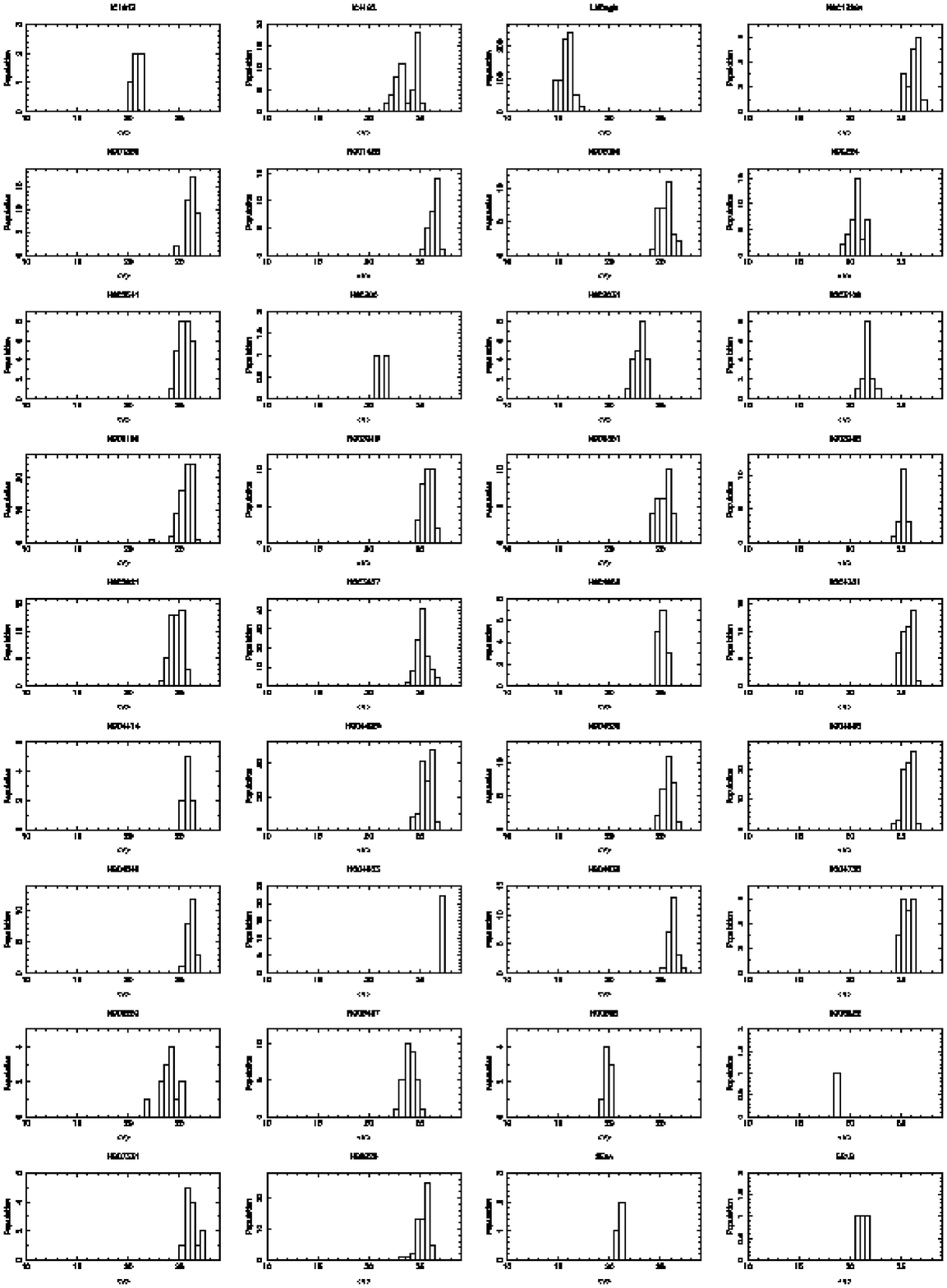}}
\caption{Histograms of apparent $\langle V \rangle $ magnitudes for each
host galaxy. On the x-axis we give $\langle V \rangle $. On the y-axis
we give the population. The completeness limit in magnitude is generally
(see text) taken from the upper limit of the most populated class.
}
\label{fig002}
\end{figure*}

\begin{figure*}[p]
\resizebox{\hsize}{!}{\includegraphics*{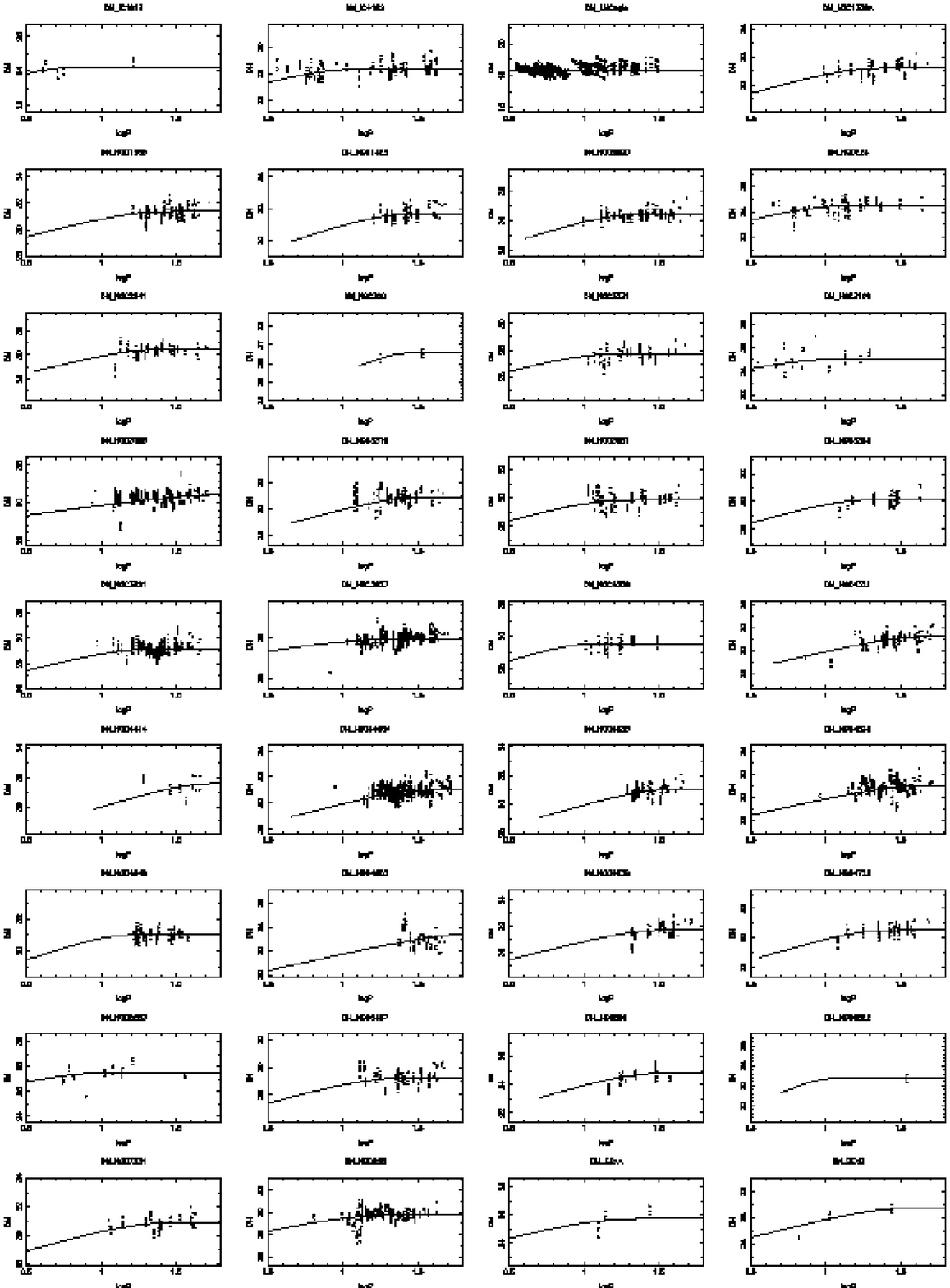}}
\caption{
Distance moduli (y-axis) from the method of sosie {\it vs.}
$\log P$ (x-axis) for each host galaxy. Each point corresponds
to an extragalactic Cepheid which is sosie of a calibrating
Cepheid. The solid curves correspond
to the adopted bias curves (section 4).
The points rejected by the cut-off at $V_{lim}$ are represented
with crosses.
}
\label{fig003}
\end{figure*}

Two additional quantities are required to apply these equations: 
\begin{itemize}
\item The limiting magnitude $V_{lim}$. 
\item The standard deviation $\sigma$ of the PL relation at
a constant $logP$.
\end{itemize}
The first quantity is derived from the histograms of $\langle V \rangle $
presented in Figure \ref{fig002} for each galaxy. We adopt for $V_{lim}$
the fainter edge of the most populated class. In a few cases
where the histogram has no dominant class, we move the value
by $\pm 0.5$ magnitude.
$V_{lim}$ values do not change significantly when one changes 
the binning size. Only one galaxy (NGC5457) changed by more than the
binning size, but its histogram shows two classes with almost the same 
population. Nevertheless, the global influence of a change in $V_{lim}$ is
discussed in section 4.3 (Table \ref{test}) and we show its
influence on each individual galaxy in Table \ref{param}.
The second quantity ($\sigma$) is derived by a direct linear regression
on each plot of Figure \ref{fig003}. The adopted quantities 
$V_{lim}$ and $\sigma$ are listed in columns 2 and 3 of Table \ref{param}.

These parameters being fixed, there is no free parameter
to adjust the bias curve to the plot of Figure \ref{fig003}
except the distance modulus $\mu$ itself which is then
determined through an iterative process.  
The final bias curves are plotted in Fig.\ref{fig003} for each host galaxy.
In column 9 of Table \ref{param} we give the number of remaining sosies 
after the cut-off at $V_{lim}$. In Fig.\ref{fig003} 
the points which are rejected by the cut-off are represented by crosses.

\subsection{Analysis of the results}
Freedman et al. (2001) recently published their final study
of their HST keyproject (HSTKP). They publish distance moduli
calculated differently to those used here.
They calibrate the PL relation with the LMC distance modulus, 
assumed to be $\mu(LMC) =18.5$. They, adopt
the V- and I-band PL relations and an extinction law giving  
${R_V}/{R_I}= 1.69$. 
In order to avoid bias, they cut their sample
at a given limiting period $\log P_l$ as explained above and they apply a
small (but still uncertain) correction for metallicity effect.

\begin{figure}[!]
\resizebox{\hsize}{!}{\includegraphics*{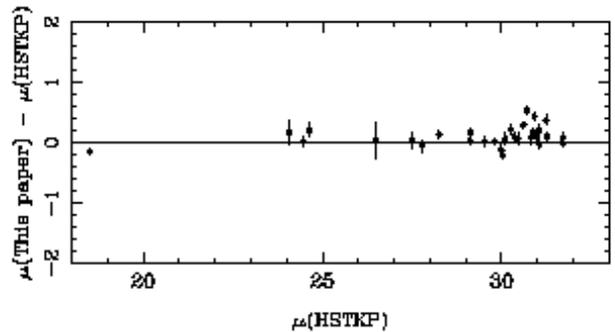}}
\caption{Comparison of the distance moduli from
Freedman et al., 2001 and those from this paper.
The general agreement is satisfactory but 
at large distances our distances become larger. 
}
\label{fig004}
\end{figure}

The comparisons between the HSTKP results and our solution is
shown in Fig.\ref{fig004} for 31 galaxies in common. There is 
a fair agreement.  
A direct regression between HSTKP distance moduli and ours 
leads to a slope which is not significantly different from one
($1.017 \pm 0.010$) and a zero point difference which is not significantly 
different from zero ($-0.11 \pm 0.16$). 
Assuming both determinations carry the same uncertainty, this means that our
distances are good within $0.16/ \sqrt{2}=0.1$ magnitude. This is the
internal uncertainty.

From a detailed check of Fig.\ref{fig004} one sees a slight departure 
from a slope of one at large distances. 
The effect is, on average, 0.17 mag. for $\mu$ larger than 30 mag.
Two possibilities can explain this discrepancy:
\begin{itemize}
\item 
The PL relation of the
GFG sample shows a departure from linearity for large $\log P$.
This effect is visible (see for instance Figure 4 in GFG) even
when one excludes the three overtone Cepheids ($\log P < 0.6$).
Judging by the error bars of individual points, this non-linearity 
seems real. 
\item 
The distance moduli of Freedman
et al. may suffer from a small residual incompleteness bias. 
Using a simulation we have shown that it is difficult to remove
the bias just by cutting the sample at a given $\log P_l$. If we refer
to our Fig. 7 in Lanoix et al. 1999a, one can see that at large distances
($\mu > 32$) the bias may reach $0.17$ magnitude after 
the $\log P$ cutoff.  At intermediate distances ($29 < \mu < 32$) 
the bias may still reach 0.08 mag.
\end{itemize}

Three external sources of uncertainty come from:
(i) the adopted ratio ${R_V}/{R_I}$, (ii) the adopted limiting
magnitude $V_{lim}$ and  (iii) from the adopted PL relation used 
for second order bias correction. 
In order to check the stability of the solution,
we repeated the previous calculations with
another PL relation (the one found by GFG for LMC),
with a variation of ${R_V}/{R_I}$ by $\pm$ 0.2 and a variation of
$V_{lim}$ by $\pm 0.5 mag$.
The results are summarized
in Table \ref{test}, where we give the mean shift between 
distance moduli from different solutions and the adopted mean distance moduli
(reference solution). 
One can see that the choice of the PL relation has no actual
influence on the result. However, a change of ${R_V}/{R_I}$
by $\pm$ 0.1 may change the mean distance modulus by nearly 0.2 magnitude
and a change of $V_{lim}$ by $0.5 mag$ may produce similar change.
The influence of $V_{lim}$ depends clearly on the actual distribution
of magnitudes. For some galaxies the effect is negligible while it is
large for some others. In order to give a better judgement of the
stability of the distance modulus with respect to the adopted $V_{lim}$,
we give the changes  $\Delta \mu^-$ when $V_{lim}$ is reduced by
0.5 mag (respectively, $\Delta \mu^+$ when $V_{lim}$ is augmented by 0.5 mag).

The actual uncertainty (internal plus external) 
can thus reach 0.3 magnitude and may be more if our actual sources of
uncertainty act in the same sense.

\begin{table}
\caption{Test of the stability of the results. We give the departure
from our reference solution for: 1) a different PL relation (note that this
PL relation is used only for the second order bias correction  2) several 
${R_V}/{R_I}$ ratios.}
\begin{tabular}{rrrrr}
\hline
$\Delta V_{lim}$& ${R_V}/{R_I}$  & $a_v$    & $b_v$    &   $\mu - \mu_{ref}$  \\
\hline
0.0             & 1.69           & $-2.769$ & $-4.063$ &    $+0.03 \pm 0.05$ \\
\hline
0.0             & 1.89           & $-3.037$ & $-4.058$ &    $-0.22 \pm 0.10$ \\
0.0             & 1.79           & $-3.037$ & $-4.058$ &    $-0.12 \pm 0.06$ \\
0.0             & 1.69           & $-3.037$ & $-4.058$ &    $0$              \\
0.0             & 1.59           & $-3.037$ & $-4.058$ &    $+0.17 \pm 0.09$ \\
0.0             & 1.49           & $-3.037$ & $-4.058$ &    $+0.45 \pm 0.21$ \\
\hline
$-0.50$         & 1.69           & $-3.037$ & $-4.058$ &    $+0.20 \pm 0.27$ \\
$-0.25$         & 1.69           & $-3.037$ & $-4.058$ &    $+0.09 \pm 0.12$ \\
$+0.25$         & 1.69           & $-3.037$ & $-4.058$ &    $-0.05 \pm 0.08$ \\
$+0.50$         & 1.69           & $-3.037$ & $-4.058$ &    $-0.08 \pm 0.10$ \\
\hline
\end{tabular}
\label{test}
\end{table}

\begin{table*}
\caption{Distance moduli calculated from this paper
using the ratio $R_V / R_I = 1.69$.
{\bf Column 1:} Name of the host galaxy.
{\bf Column 2:} The adopted limiting magnitude $V_{lim}$.
{\bf Columns 3:} Standard deviation $\sigma$. 
{\bf Column 4:} The adopted distance modulus and its mean error. An asterisk
marks the distance moduli of galaxies having nearly a Solar metallicity.
{\bf Column 5:} The change $\Delta \mu^-$ of the distance modulus 
when $V_{lim}$ is reduced by 0.5 mag (i.e., a brighter limit).
{\bf Column 6:} The change $\Delta \mu^+$ of the distance modulus 
when $V_{lim}$ is augmented by 0.5 mag (i.e., a fainter limit).
{\bf Column 7:} The number of sosie Cepheids after the $V_{lim}$ cut-off.
}
\begin{tabular}{lrrrrrr}
\hline
galaxy & $V_{lim}$ & $\sigma$ &  $\mu \pm m.e.$ & $\Delta \mu^-$&$\Delta \mu^+$  &n\\
\hline
IC1613  & 21.5& 0.36&   $ 24.23\pm  0.19$ & 0.16&-0.09&  12\\
IC4182  & 25.0& 0.50&   $ 28.39\pm  0.07$ & 0.00&-0.05& 169\\
LMCogle & 16.5& 0.24&   $ 18.36\pm  0.03$ & 0.03& 0.01& 947\\
NGC1326A& 27.0& 0.46&   $ 31.24\pm  0.09$ & 0.07&-0.07&  70\\
NGC1365 & 27.0& 0.43& * $ 31.38\pm  0.07$ & 0.04&-0.03& 152\\
NGC1425 & 27.0& 0.35& * $ 31.70\pm  0.06$ & 0.31&-0.08&  99\\
NGC2090 & 26.0& 0.31& * $ 30.44\pm  0.07$ & 0.09&-0.03& 103\\
NGC224  & 21.0& 0.47& * $ 24.50\pm  0.08$ & 0.19&-0.08& 106\\
NGC2541 & 26.0& 0.35& * $ 30.47\pm  0.07$ & 0.03&-0.07&  88\\
NGC300  & 21.5& 0.14&   $ 26.54\pm  0.29$ & 0.08&-0.11&   4\\
NGC3031 & 24.0& 0.47& * $ 27.75\pm  0.10$ & 0.09&-0.05&  92\\
NGC3109 & 22.0& 0.61&   $ 25.10\pm  0.16$ & 0.30& 0.11&  31\\
NGC3198 & 26.0& 0.86&   $ 31.23\pm  0.07$ & 0.70&-0.17& 187\\
NGC3319 & 26.0& 0.38&   $ 30.91\pm  0.06$ & 0.89&-0.03&  88\\
NGC3351 & 26.0& 0.50& * $ 29.88\pm  0.08$ & 0.07& 0.01& 110\\
NGC3368 & 26.0& 0.39& * $ 30.17\pm  0.10$ & 0.09&-0.05&  74\\
NGC3621 & 25.0& 0.43& * $ 29.15\pm  0.07$ & 0.06& 0.04& 152\\
NGC3627 & 26.0& 0.66& * $ 29.80\pm  0.06$ & 0.06&-0.03& 369\\
NGC4258 & 26.0& 0.34& * $ 29.53\pm  0.10$ & 0.09&-0.01&  65\\
NGC4321 & 26.0& 0.47&   $ 31.35\pm  0.06$ & 0.85&-0.28&  78\\
NGC4414 & 26.0& 0.33&   $ 31.62\pm  0.09$ & 0.78&-0.15&  18\\
NGC4496A& 26.0& 0.41&   $ 31.03\pm  0.04$ & 0.42&-0.06& 280\\
NGC4535 & 26.0& 0.38& * $ 31.08\pm  0.07$ & 0.15&-0.08&  64\\
NGC4536 & 26.0& 0.52& * $ 31.04\pm  0.06$ & 0.18&-0.12& 153\\
NGC4548 & 27.0& 0.31& * $ 31.03\pm  0.08$ & 0.05&-0.01& 100\\
NGC4603 & 28.0& 0.84&   $ 33.70\pm  0.09$ & 0.00&-0.52&  79\\
NGC4639 & 27.0& 0.52&   $ 31.80\pm  0.08$ & 0.41&-0.12&  77\\
NGC4725 & 26.0& 0.36& * $ 30.53\pm  0.11$ &-0.01&-0.04&  53\\
NGC5253 & 24.5& 0.52& * $ 27.53\pm  0.14$ & 0.13&-0.01&  30\\
NGC5457 & 25.0& 0.51& * $ 29.30\pm  0.07$ & 0.10&-0.01& 102\\
NGC598  & 20.0& 0.34&   $ 24.83\pm  0.12$ &-0.38&-0.23&  22\\
NGC6822 & 19.5& 0.14&   $ 23.38\pm  0.52$ & 0.00& 0.00&   4\\
NGC7331 & 26.5& 0.50&   $ 30.93\pm  0.12$ & 0.39&-0.05&  48\\
NGC925  & 26.0& 0.62& * $ 29.83\pm  0.06$ & 0.02&-0.08& 238\\
SEXA    & 22.0& 0.62&   $ 25.75\pm  0.23$ & 0.31&-0.15&  14\\
SEXB    & 22.0& 0.51&   $ 26.77\pm  0.18$ & 0.53&-0.27&   9\\
\hline
\end{tabular}
\label{param}
\end{table*}

\section{Conclusion}
The distance scale can be calibrated using galactic Cepheids.
LMC provides us with numerous Cepheids located at the same distance. 
This gives a way to derive an accurate slope for the Cepheid
PL relation. 
But its low metallicity (with respect to most of the galaxies
of the sample) is a cause of suspicion;
we are not sure that this slope can be applied to all kinds of 
metallicity.

So, we preferred, in a first step, 
to calibrate the distance scale
by using accurate distances of galactic Cepheids published by
Gieren, Fouqu\'e and Gomez (1998). These distances are based
on the geometrical Barnes-Evans method. 

Further, we applied the concept of 'sosie' (Paturel, 1984) to extend 
distance determinations to extragalactic Cepheids without having to 
know either the slope or the zero-point of the PL relation. 
The distance moduli are obtained in a straightforward way.
For the calibrating galactic Cepheids we 
checked the internal coherence from the same method.

The correction for the extinction is made by using two bands 
($V$ and $I$) according to the principles introduced by 
Freedman and Madore (1990). There is no need for color excess estimation.

Finally, the incompleteness bias is corrected according to the precepts
introduced by Teerikorpi (1987). Without any free parameters (except
the distance modulus itself), the bias curve calculated for each 
individual host galaxy fits very well the observed distance moduli.
This gives us confidence in our final distance moduli.
Nevertheless, the small departure from the measurements published
recently by Freedman et al. (2001) at distances larger than 10Mpc 
($\mu=30$) must be clarified. 

In order to bypass the uncertainty due to metallicity effects it
is suggested to consider only galaxies having nearly the same
metallicity as the calibrating Cepheids (i.e. Solar metallicity).
In Table \ref{param} the distance moduli that can be considered
as more secure are noted with an asterisk ($\ast$). Galaxies 
with $\Delta \mu$ larger than $\approx 0.3$ mag. or 
with small $n$ do not receive this flag.
For a given ratio ${R_V}/{R_I}$, the uncertainty of the distances
is about 0.1 magnitude but the total uncertainty may be about 0.3
magnitude.
The choice of a given ${R_V}/{R_I}$ ratio is a first source of
uncertainty. The actual
ratio depends on the extinction law in our Galaxy, on the extinction law
in the host galaxy and on the color of the considered Cepheid. 
For the future it would be interesting to search for a clue allowing 
us to decide which value is the best in a given direction for a Cepheid in
a given host galaxy. 
The proper determination of the limiting magnitude of the sample 
is a second source of uncertainty. It can be accurately determined
only when a large number of Cepheids is available to provide us
with good statistics.

Presently, the calibration of the distance scale can barely be
better than $\sigma_{\mu}=0.3$ magnitude. Thus, the uncertainty 
on the Hubble constant, $\sigma(H) \approx \sigma_{\mu}  H / 5 \log{e}$, 
cannot be better than  about $10 km.s^{-1}.Mpc^{-1}$.

\begin{appendix}
\section{The extragalactic Cepheid database}
The description of this database was given by Lanoix et al. (1999b).
Because the database is no longer available on the world-wide-web
the present data are published in electronic form in the A\&A archives
at CDS. All the data are made available, even when they are not used in 
the present paper, where only Normal Cepheids in V and I-bands are considered. 
Additional measurements were collected including 
the LMC ones by Udalski et al. (1999)\footnote{Note that we kept 720 normal
Cepheids among the 1182 available with $\log P > 0.5$.} and those
by Gibson et al. (1998, 1999). Data are now available
for 2449 Cepheids of 46 galaxies (instead of 1061 Cepheids of 33 galaxies).

The identification of a Cepheid is given on a first line as follows:
\begin{itemize}
\item the name of the host galaxy, 
\item the name of the Cepheid and the bibliographic code from where this name 
is taken,
\item the adopted period (in log)
\item The classification of the shape of the light curve, 
following Lanoix et al. (1999b)
\end{itemize}
On this first line we also give the number of measurements attached to this 
Cepheid. 
Note that the Cepheid name for LMC is simply the Cepheid number from 
Udalski et al., without the field number (SC), that was not needed here
(only three Cepheids appear with
the same number in different fields: 1, 16 and 19, but they are not in our 
list).
We tried to keep the Cepheid name of the first discovery. This was not always
done, e.g., the names given by Graham (1984) are referenced as Mad87 because of 
the renumbering adopted by Madore (1987).

On the following lines, individual measurements are given: 
\begin{itemize}
\item the magnitude,
\item the type of magnitude (mean, maximum, minimum, average) coded according 
to Lanoix et al.,
\item the photometric bands (B,V,R,I ...) coded according to Lanoix et al.(1999b),
\item the reference code. The full reference and the associated code appears 
in the references.
\end{itemize}

A sample is given below to show how the data are organized.

\begin{verbatim}



IC1613  V1         Fr88a  0.7480     N    8
 21.36     mea       B   Fr88a
 20.79     mea       V   Fr88a
 20.36     mea       R   Fr88a
 20.14     mea       I   Fr88a
 20.50     max       B   Sa88a
 22.03     min       B   Sa88a
 21.27     ave       B   Sa88a
 21.39     mea       B   Sa88a
IC1613  V20        Fr88a  1.6220     B    5
 16.66               H   Ala84
 18.98     max       B   Sa88a
 20.71     min       B   Sa88a
 19.85     ave       B   Sa88a
 19.90     mea       B   Sa88a
IC1613  V22        Fr88a  2.1650     S    9
 15.47               H   Ala84
 19.10     mea       B   Fr88a
 17.75     mea       V   Fr88a
 17.14     mea       R   Fr88a
 16.62     mea       I   Fr88a
 17.74     max       B   Sa88a
 20.44     min       B   Sa88a
 19.09     ave       B   Sa88a
 19.09     mea       B   Sa88a
IC1613  V25        Fr88a  0.9600     B+   5
 18.62               H   Ala84
 20.10     max       B   Sa88a
 21.84     min       B   Sa88a
 20.97     ave       B   Sa88a
 20.87     mea       B   Sa88a
IC1613  V53        Fr88a  0.5900     O    3
 21.13     max       B   Car90
 21.70     min       B   Car90
 21.46     mea       B   Car90
...
...
...
\end{verbatim}
\end{appendix}

\acknowledgements{We thank the HST teams for making their data
available in the literature prior to the end of the project. 
We thank R. Garnier, J. Rousseau  and P. Lanoix for having
participated to the maintenance of our Cepheid database.
We thank P. Teerikorpi for his comments and the anonymous referee
for very constructive remarks.
}


\end{document}